\documentclass[12pt,a4paper]{article}

\usepackage{amsmath,amssymb,amsfonts}
\usepackage[dvips]{graphicx}
\usepackage{epsfig}
\usepackage{cite}
\usepackage{amsmath,amssymb,amsfonts,graphicx}
\usepackage{epsfig}

\newcommand{\beq}{\begin{eqnarray}}
\newcommand{\eeq}{\end{eqnarray}}
\newcommand{\bea}{\begin{eqnarray}}
\newcommand{\eea}{\end{eqnarray}}

\newcommand{\bec}{\begin{center}}
\newcommand{\eec}{\end{center}}

\def\N{{\cal N}}
\def\D{\cal D}

\numberwithin{equation}{section}

\begin{document}

\title{
\begin{flushright}\ \vskip -1.5cm {\small {IFUP-TH}}\end{flushright}
\vskip 30pt
\bf{Baby Skyrme Model, Near-BPS Approximations and Supersymmetric Extensions}
\vskip 15pt}
\author{
S. Bolognesi$^{(1)}$ and W. Zakrzewski$^{(2)}$ \\[20pt]
{\em \normalsize  
$^{(1)}$Department of Physics “E. Fermi” and INFN, University of Pisa}\\[0pt] 
{\em \normalsize  
Largo Pontecorvo, 3, Ed. C, 56127 Pisa, Italy } \\
{\normalsize Email: stefanobolo@gmail.com}\\[5pt] 
{\em \normalsize  
$^{(2)}$Department of Mathematical Sciences,}\\[0pt] 
{\em \normalsize Durham University, Durham DH1 3LE, U.K.}\\
{ \normalsize   Email: w.j.zakrzewski@durham.ac.uk}\\ [5pt] 
}
\vskip 10pt
\date{September 2014}
\maketitle
\begin{abstract}
We study the baby Skyrme model as a theory that interpolates between two distinct BPS systems. 
For this a near-BPS approximation can be used when there is a small deviation from each of the two BPS limits. 
We provide analytical explanation and numerical support for the validity of this approximation.
We then study the set of all possible supersymmetric extensions of the baby Skyrme model with $\N=1$ and the particular ones with extended $\N=2$ supersymmetries and relate this to the above mentioned almost-BPS approximation.
\end{abstract}
\newpage

\section{Introduction}

The baby Skyrme model  in $(2+1)$ dimensions \cite{Leese:1989gi,Piette:1994ug} has been widely investigated, both for its own interests and for being a toy model more sophisticated theories in higher dimensions.   
In this paper we focus our attention on its features as a theory that  interpolates between two distinct BPS systems \cite{Adam:2010jr,Gisiger:1996vb}. 
We note that, after a convenient rescaling, the model depends on only one parameter $\zeta$ that can be set to take values in the interval $0 \leq \zeta \leq 1$.  At the edges of this interval there are two distinct BPS models: the $O(3)$ sigma model at $\zeta =0$, and the restricted  baby Skyrme model at $\zeta =1$. 
Near both edges of this interval an almost-BPS approximation can be used to obtain an analytic approximation of the soliton solution. The exact solution, which we obtain numerically for the first topological sector, flows to this approximation as the parameter $\zeta$ goes to $0$ or to $1$.

An analogue of this near-BPS approximation has been used before in the context of holographic QCD \cite{Bolognesi:2013nja,Hong:2007kx,Hata:2007mb,Bolognesi:2013jba,Bolognesi:2014dja} and generalized Skyrme model \cite{Adam:2010ds,Adam:2013tda,Adam:2014xfa,Speight:2014fqa}. In both cases, one of the physical motives is to have a model that reproduces the small binding energies observed in nuclear physics. So the near-BPS approximation has both mathematical and phenomenological interest.

The  baby Skyrme model possesses various supersymmetric extensions which all have in common the same bosonic sector. These supersymmetric extensions of the baby Skyrme model were  first discussed in  \cite{Adam:2011hj,Adam:2013awa}, following earlier attempts to supersymmetrize Skyrme-like theories in $3+1$ dimensions \cite{Bergshoeff:1984wb,Freyhult:2003zb} (see also more recently \cite{Nitta:2014pwa}). 
These supersymmetric theories are in general ${\cal N}=1$ supersymmetric, thus with two real supercharges,   and become  ${\cal N}=2$  at the two ends of the interval.   The almost-BPS properties of the almost-BPS theory can then be understood in terms of the quantum supersymmetry algebra.

The paper is organized as follows: In Section \ref{first} we study the bosonic baby Skyrme model and its near-BPS limits for various choices of potentials. In Section \ref{firstsusy} we consider the $\N=1$ supersymmetric extensions of these theories.
In Section \ref{secondsusy} we study the $\N=2$ extensions and their BPS properties. We conclude in Section \ref{conclusion} with some open questions.

\section{The bosonic baby Skyrme model}
\label{first}

The action for the $O(3)=S^2$ baby Skyrme model is
\beq
\label{actionbsm}
S = \int d^3x \left( \frac{\theta_2}{2} \partial_{\mu} \vec{\phi} \cdot  \partial^{\mu} \vec{\phi} - \frac{\theta_4}{2}( \partial_{\mu} \vec{\phi} \times \partial_{\nu} \vec{\phi} )\cdot  ( \partial^{\mu} \vec{\phi} \times \partial^{\nu} \vec{\phi} ) - \theta_0 V(\vec{\phi}) \right) \ ,
\eeq
with the target space subject to the constraint  $\vec{\phi} \cdot \vec{\phi}=1$. 
We consider this model for a class of  potentials of the following form
\beq
\label{pot}
V(\vec{\phi}) =  \left(\frac{1-\hat{n}\cdot\vec{\phi}}{2} \right)^k \ ,
\eeq
where $\hat{n}$ is a unit vector and $k$ an integer. This family of potentials contains, for example, the old baby Skyrme model for $k=1$ and  the so called holomorphic model for $k=4$ \cite{Leese:1989gi}. 
In addition to the arbitrariness of the functional form of the potential  we have, in general, three parameters $\theta_{0,2,4}$ in the model. Rescaling the action by an overall constant, and rescaling the length scale,  we can effectively reduce this arbitrariness to  having only a one parameter family of Lagrangians.  

 Shortly we choose a parametrization which is the most  convenient for us; namely to describe the flow between two BPS systems that we want to study in this paper.

The first BPS system is the pure sigma model whose Lagrangian is given by
\beq
{\cal L}_2 =   \frac{1}{2} \partial_{\mu} \vec{\phi} \cdot  \partial^{\mu} \vec{\phi} \ , 
\eeq
which, in  the $CP(1)$ formulation, takes the form 
\beq
\label{L2w}
{\cal L}_2 =   \frac{1}{(1+|w|^2)^2} \partial_{\mu} w \partial^{\mu} \bar{w}  \ .
\eeq
This model has a BPS bound which is saturated by the holomorphic and anti-holomorphic solutions 
\beq
\label{bound1}
E_{BPS} = 4 \pi |Q| \ ,
\eeq
where $Q$ is the topological charge.

The second BPS system is the so called restricted baby Skyrme model and its Lagrangian consists of 
only two terms, the term with quartic derivatives and the  potential term: 
\beq
{\cal L}_{4,0} =   - \frac{1}{2}( \partial_{\mu} \vec{\phi} \times \partial_{\nu} \vec{\phi} )\cdot  ( \partial^{\mu} \vec{\phi} \times \partial^{\nu} \vec{\phi} ) -  V(\vec{\phi}) \ .
\eeq
For the potential of the form (\ref{pot}), and using  the $CP(1)$ formulation, the Lagrangian of the restricted baby Skyrme model is  described by
\beq
\label{L40w}
{\cal L}_{4,0} =  \frac{1}{(1+|w|^2)^4}  \partial_{\mu}w   \partial^{\nu}  \bar{w} (\partial_{\mu}w   \partial^{\nu} \bar{w} -  \partial_{\nu}w   \partial^{\mu} \bar{w}) \phantom{\frac{1}{2}} - \frac{|w|^{2k}}{\left(1+|w|^2\right)^k} \ .
\eeq
This model also  has also a BPS bound and its solutions satisfy:
\beq
\label{bound2}
E_{BPS} = \frac{8 \pi }{k+2} |Q| \ .
\eeq

The full baby Skyrme model can be thought of as  an interpolation between these two BPS systems.
By rescaling the action and the length scale, we can choose the parameters in (\ref{actionbsm}) to be of the form
\beq
\theta_2 = 1- \zeta \ , \qquad  \qquad 
\theta_0 = \theta_4 = \frac{ \zeta (k+2)}{2}  \ ,
\eeq
and so the full Lagrangian can be written as 
\beq
{\cal L} =  (1-\zeta) {\cal L}_{2} + \frac{ \zeta (k+2)}{2} {\cal L}_{4,0} \ ,
\eeq 
where $ {\cal L}_{2}$ and ${\cal L}_{4,0}$ are given in (\ref{L2w}) and (\ref{L40w}).
The  parameter $\zeta$ takes value in an interval $[0,1]$ and the boundaries of the interval represent the two BPS systems.
Note that for this choice of parameters the total bound, which is the sum of the two BPS bounds (\ref{bound1}) and (\ref{bound2}), is fixed to be $ 4 \pi |Q|$ for every value of $\zeta$.
The existence of this bound for the full system follows directly from the existence of the  two bounds of the two BPS systems taken in isolation  \cite{Adam:2010jr}.  In general, the bound can be saturated only at the edges of the interval as we shall demonstrate below.

To find a one soliton solution we consider the radial ansatz
\beq
w(r,\theta) = e^{i \theta} f(r)
\eeq
for which the profile function $f(r)$ has  to satisfy the boundary conditions $f(r\to 0)= \infty$ and $f(r\to \infty)=0$. 
The energy functional in terms of $f(r)$ is now given by 
\bea
\frac{E}{4\pi} &=& \int dr \left\{  \frac{r(1-\zeta)}{(1+f^2)^2} \left( f'^2 + \frac{f^2}{r^2}\right) \right. \nonumber \\ && \ \  \qquad \left. + \frac{\zeta(k+2)}{4} \left( \frac{4 f'^2 f^2}{r (1+f^2)^4} + \frac{r f^{2k} }{(1+f^2)^k} \right) \right\} \ .
\label{aaa}
\eea
The exact forms of this profile function can be obtained by minimizing this functional for various values of $\zeta$. The profile functions, for all values of $\zeta$,  always diverge like $f(r) \simeq \lambda/r$  as $r \to 0$. To find the profile function numerically we can use the `shooting method' {\it i.e.} varying the parameter $\lambda$ until we find that the other boundary condition (at infinity) is also satisfied.

Next we consider a near-BPS approximation to describe the soliton solutions near the two edges of the interval. We first describe our approach in detail for the first edge, $\zeta \to 0$, {\it i.e.} the one close to the pure sigma model. This method is very similar to the one discussed in \cite{Bolognesi:2013jba} for a holographic model in which the role of the potential was played by the space-time curvature.  Earlier uses of this method for different theories can be found in \cite{Bolognesi:2013nja,Hong:2007kx,Hata:2007mb,Bolognesi:2013jba,Adam:2010ds,Adam:2013tda,Speight:2014fqa}. 
 A similar, but not equivalent, approach for the study of the baby Skyrme model can also be found in \cite{Ioannidou:2002cq}.

A solution of the one soliton profile of the pure sigma model ${\cal L}_2$ can be taken in the form of  holomorphic function
\beq
\label{holoansatz}
f(r) = \frac{\lambda}{r} \ ,
\eeq
where $\lambda$ describes the scale of the lump, and is a free parameter. We put this ansatz into 
(\ref{aaa}) and determine the value of $\lambda$ that minimize the total energy. The result of the minimization gives us
\beq
\label{lambdamin}
\lambda_* = \frac{2^{1/2}(k-1)^{1/4}}{3^{1/4}} \ .
\eeq
Note that this approach can be used only if ${\cal L}_{4,0}$, evaluated on the holomorphic ansatz, is convergent. This is true for $k>1$ and thus  excludes the old baby Skyrme model which we will discuss separately.
The total energy for the holomorphic ansatz, evaluated for the minimum (\ref{lambdamin}), is  then
\beq
\label{firstorderE}
E = 4 \pi + 4 \pi \zeta \left( \frac{k+2}{(k-1)^{1/2}2 \sqrt{3}} -1  \right) \ .
\eeq

This result can now be used in two different ways. First of all it provides an upper-bound to the exact soliton energy, which is valid for any value of $\zeta$.  Secondly, in the limit $\zeta \to 0$, the exact solutions become well approximated by the holomorphic ansatz (\ref{holoansatz}) at the scale (\ref{lambdamin}), and  (\ref{firstorderE}) gives the correct first order expansion of the soliton mass near $\zeta =0$. Later we will present an analytic argument explaining why this approximation can be trusted and we will present numerical evidence for this claim.

The other BPS approximation is very analogous in spirit. To discuss it we first consider any solution of the restricted baby Skyrme model ${\cal L}_{4,0}$. Such a solution would strongly depend on the value of the parameter $k$, so for the moment we consider $k=2$. 
The restricted baby Skyrme model has an infinitely large space of solutions, due to its area-preserving diffeomorphism invariance. The solution with radial symmetry is of the form
\beq
\label{otherbpsf}
f(r) =  \frac{1}{\sqrt{e^{r^2/2}-1}} \ . 
\eeq
Minimizing ${\cal L}_{2}$ on the space of solutions is quite simple for the one-soliton case; we just pick the radially symmetric solution. The energy of this solutions is then given by:
\beq
\label{firstorderotherbpsE}
E = 4 \pi + 4 \pi (1-\zeta) \left( \frac{\pi^2}{12} + \frac{\log{2}}{2} -1\right) \ .
\eeq

Again this result has a double interpretation. It is either an exact upper bound, which can be taken together with (\ref{firstorderE}), or it is an approximate solution valid near $\zeta =1$.

Another case we will consider explicitly is  the $k=6$ case for which the  solution  with radial symmetry of ${\cal L}_{4,0}$
is given by
\beq
f(r) = \sqrt{\frac{1}{\sqrt{1+r^2}-1}} \ ,
\eeq
and  its energy  is 
\beq
\label{firstorderotherbpsEk6}
E = 4 \pi + 4 \pi (1-\zeta) \left(\log{2} - \frac{5}{8}\right) \ .
\eeq

\begin{figure}[ht!]
\centerline{
\epsfxsize=6.5cm \epsfbox{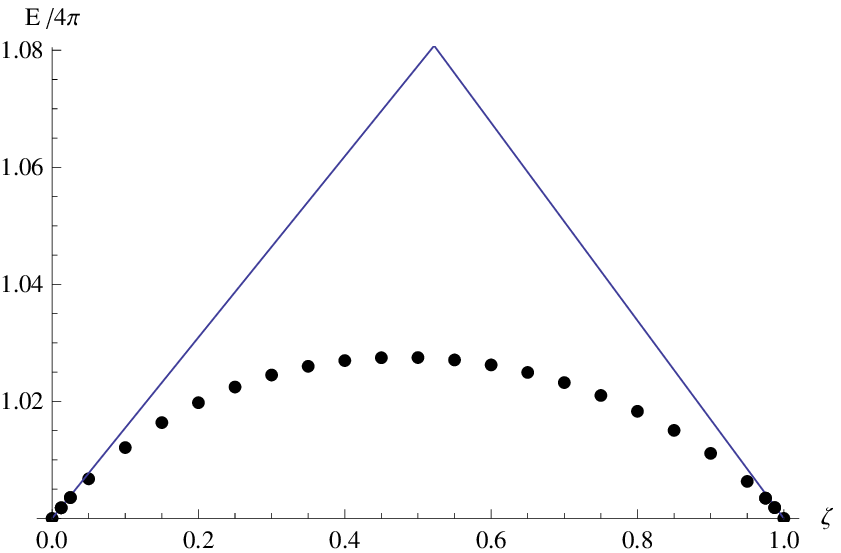} \qquad 
\epsfxsize=6.5cm \epsfbox{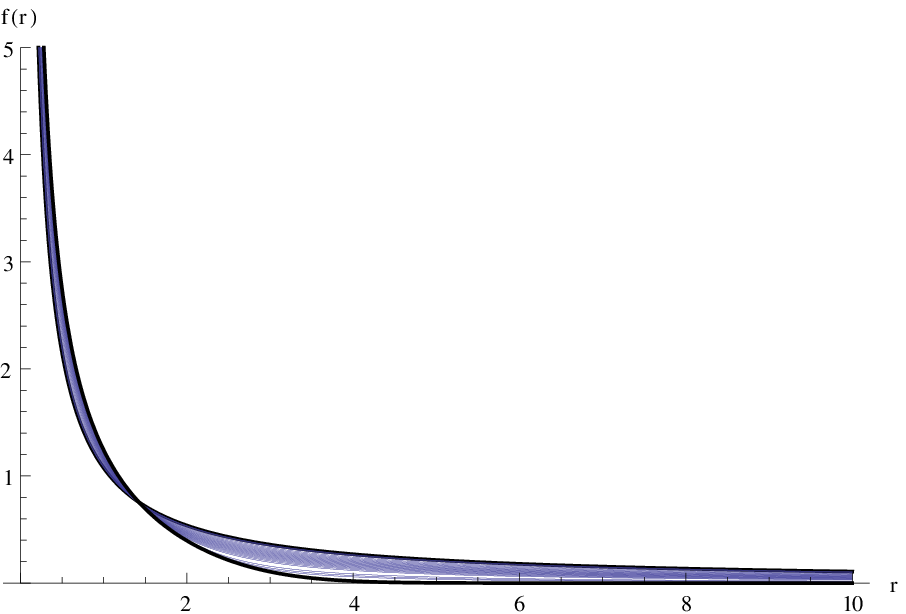}}
\centerline{
\epsfxsize=6.5cm \epsfbox{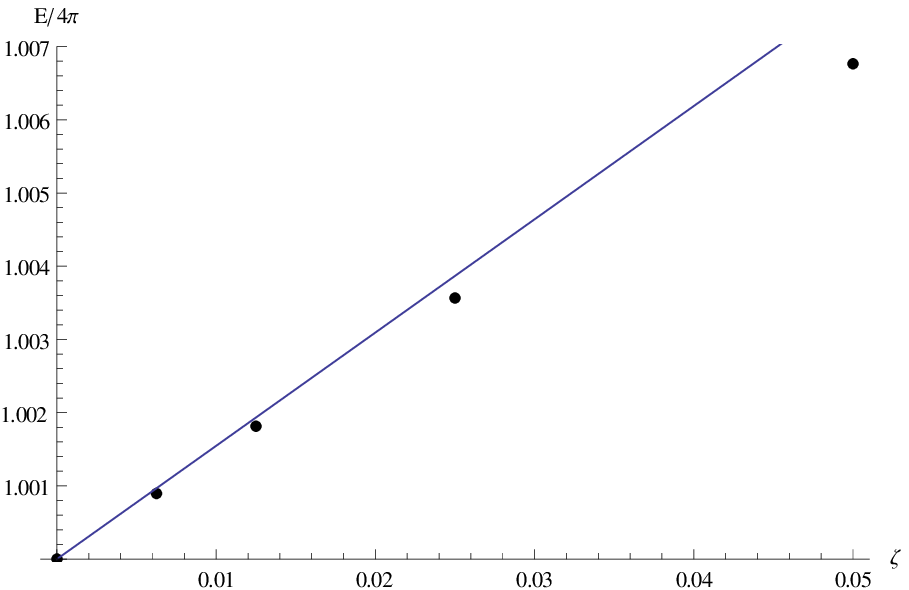} \qquad 
\epsfxsize=6.5cm \epsfbox{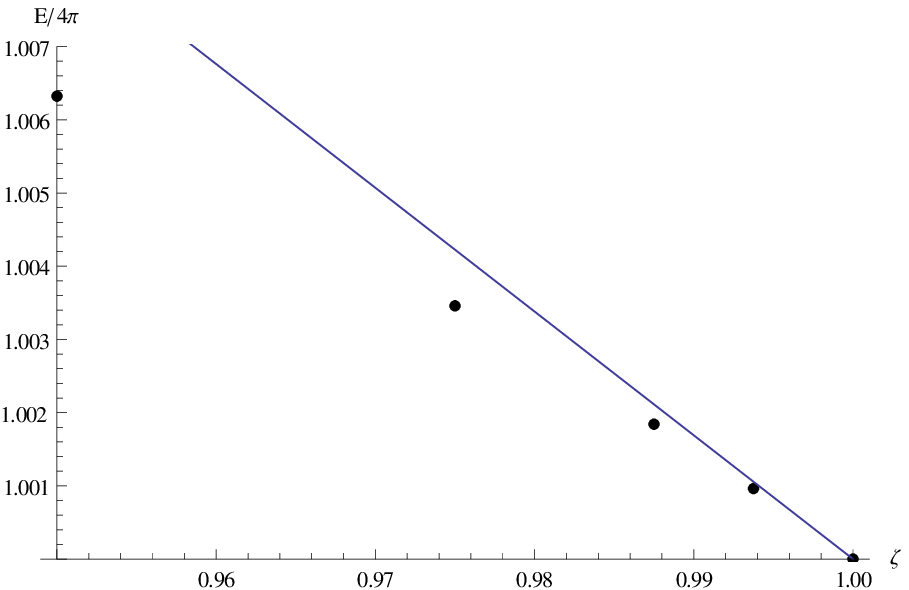}}
\caption{{\footnotesize In the first plot first row, the mass of the soliton normalized to the BPS lower bound $4\pi$, for $k=2$ is plotted as a function of $\zeta$. The upper bounds are the two near-BPS approximations (blown up in the plots below). The second plot first row presents the corresponding radial profiles for $f(r)$ for various
values of $\zeta$. Thus it shows the flow between the two almost-BPS solutions as $\zeta$ varies over the interval $[0,1]$. the plots in the second row are the mass plot zoomed near the two edges of the interval.}}
\label{flowk=2}
\end{figure}

\begin{figure}[ht!]
\centerline{
\epsfxsize=6.5cm \epsfbox{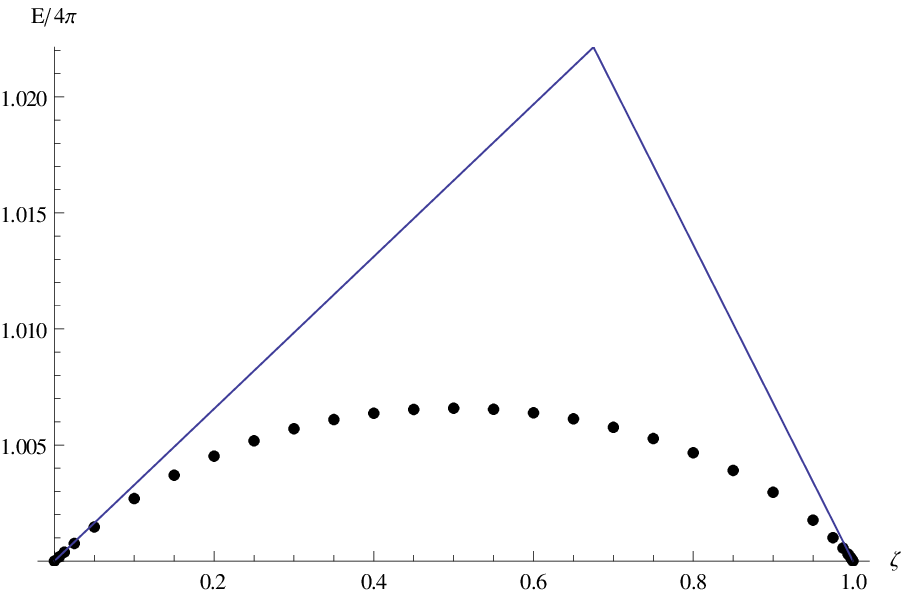} \qquad 
\epsfxsize=6.5cm \epsfbox{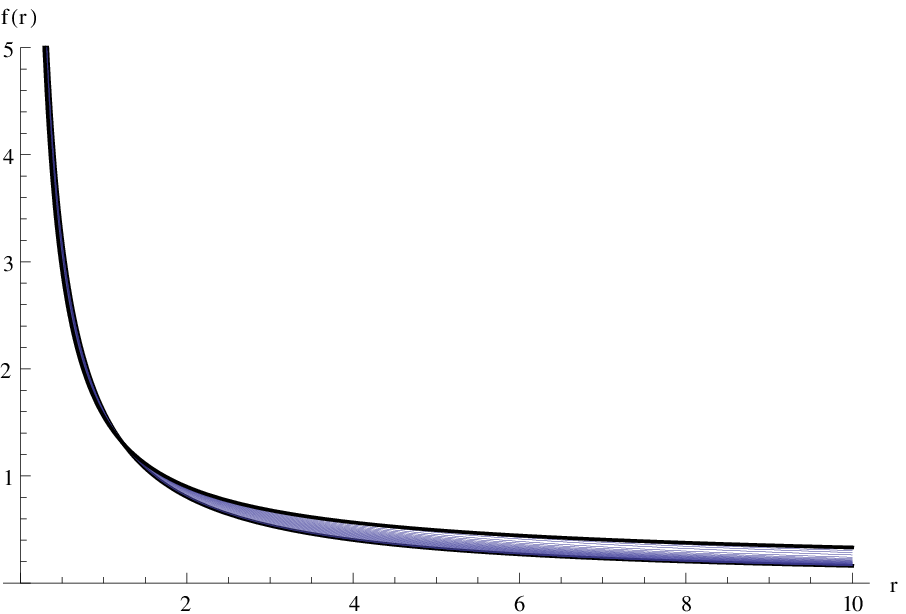}}
\centerline{
\epsfxsize=6.5cm \epsfbox{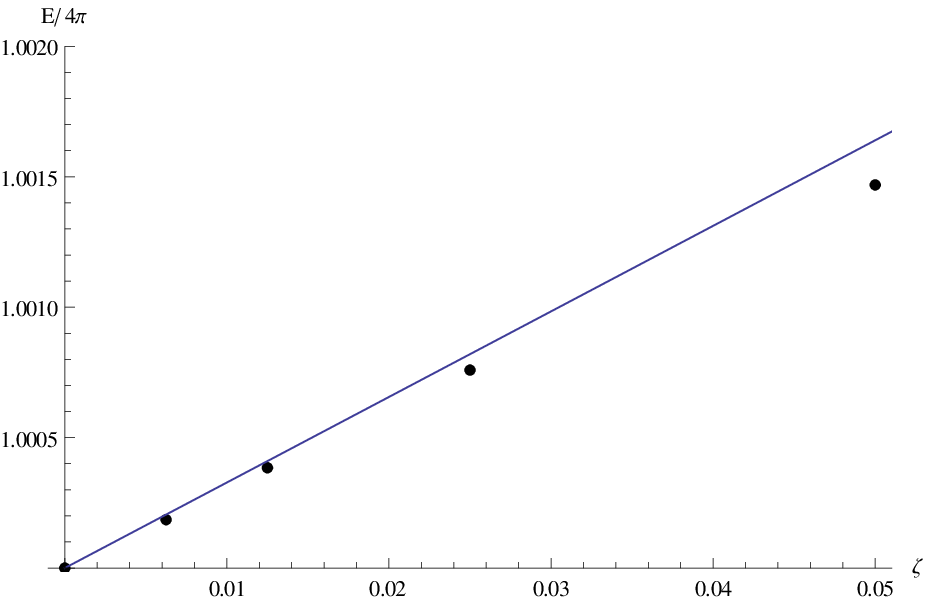} \qquad 
\epsfxsize=6.5cm \epsfbox{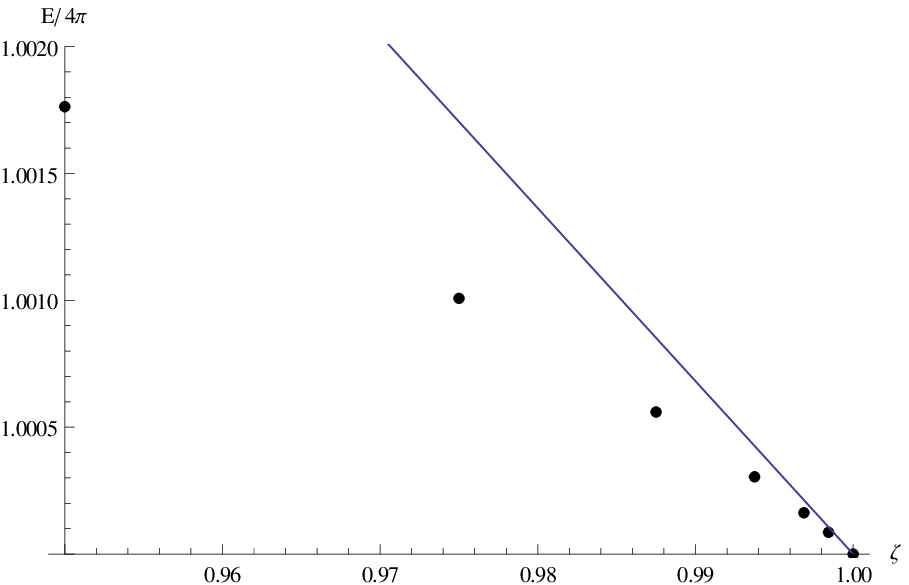}}
\caption{{\footnotesize As Figure \ref{flowk=2} but for $k=6$. }}
\label{flowk=6}
\end{figure}

We can determine the profile functions of the one-soliton fields numerically for $k=2,6$ and for various values of $\zeta$. Our results are presented Figures \ref{flowk=2} and \ref{flowk=6}.  The first plot in the first row in both sets of figures  is of  the soliton mass, normalized to $4\pi$, and compared with the various bounds. The plots show that the lower BPS bound and the two upper bounds (\ref{firstorderE}) and (\ref{firstorderotherbpsE}) form a triangle.  The energy near the two edges is well approximated by the upper bounds.
The second plot in the first row in the figures show the corresponding profile functions $f(r)$
for different values of $\zeta$. It is quite clear that the full functions converge to the BPS solutions near the two edges. In the second row of the figures we zoom near the two edges of the mass plot to show that the linear expansion is well captured by the near-BPS ansatz.

The case $k=4$ is special. This is the case of the holomorphic potential for which a holomorphic solution for the charge one sector exists for all values of $\zeta$. This is due to the fact that the moduli space for ${\cal L}_{2} $ and the moduli space for ${\cal L}_{4,0}$ intersect at one point. In this case there is no flow and the total BPS bound is always saturated.

\begin{figure}[ht!]
\centerline{
\epsfxsize=6.5cm \epsfbox{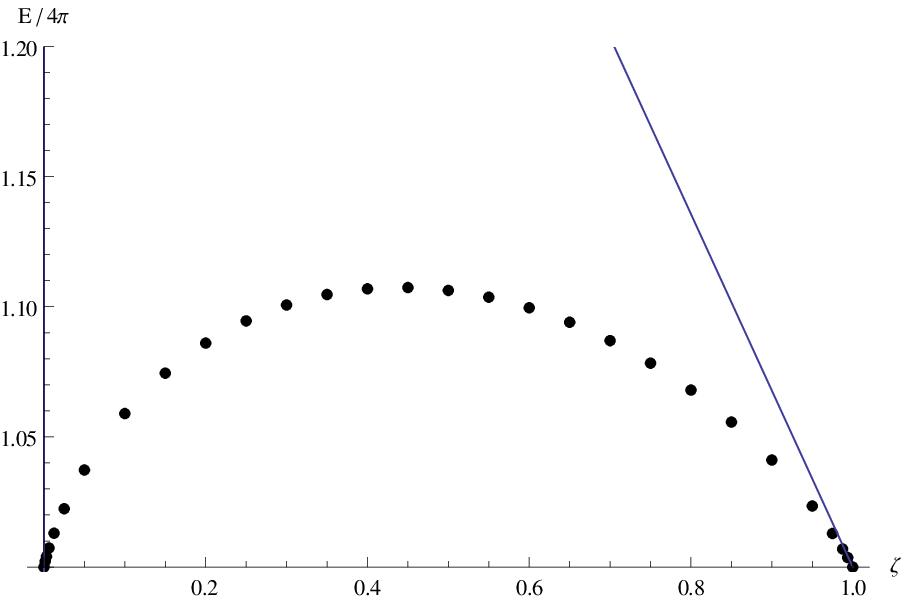} \qquad 
\epsfxsize=6.5cm \epsfbox{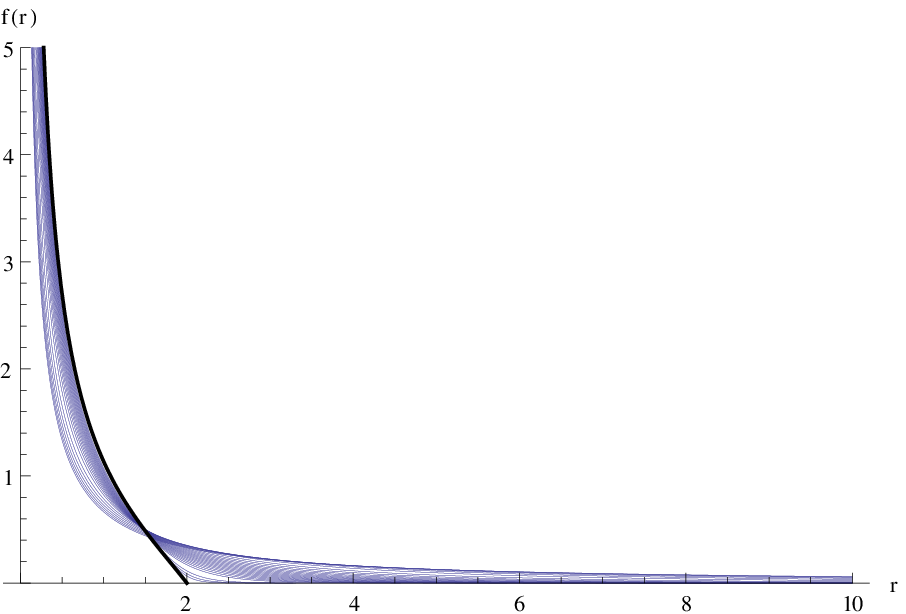}}
\centerline{
\epsfxsize=6.5cm \epsfbox{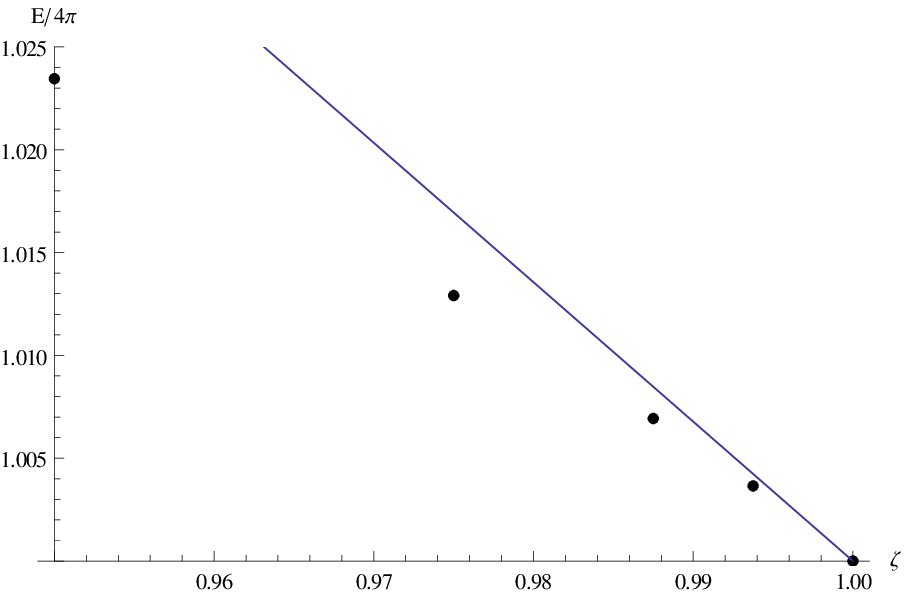}}
\caption{{\footnotesize As Figures \ref{flowk=2} and \ref{flowk=6} but this time  for $k=1$.   }}
\label{flowk=1}
\end{figure}

So we see that for the cases $k > 1$ everything works very well.
The only exception is, as stated before, the case of $k=1$ near the first edge of the interval $\eta \to 0$. 
Clearly, when $k=1$ we cannot use the equation (\ref{firstorderE}). The reason for this is that the holomorphic solution (\ref{holoansatz}) diverges when evaluated on ${\cal L}_{4,0}$. The only information we can extract from this  analysis is that the solution converges to a singular holomorphic function as $\lambda_* \to 0$ and the derivative of the energy with respect to $\zeta$ is infinite at $\zeta =0$. 
For the other edge we can still use the near BPS approximation. The solution at $\zeta =1$ is given by the following  function (with compact support):


\beq
 f(r) = \left\{ \begin{array}{ll} \frac{4-r^2}{r \sqrt{8-r^2}} &  r \leq 2 \\
0 & r\geq 2 \end{array}
 \right. 
\eeq

The  energy evaluated for this function is given by
\beq
\label{firstorderotherbpsEk1}
E = 4 \pi + 4 \pi (1-\zeta) \left(2\log{2} - \frac{17}{24}\right) \ .
\eeq
All this is confirmed by the numerical calculations, the results of which are presented in Figure \ref{flowk=1}.

Now we present an explanation of why the  near-BPS approximation works in a general case, using a finite dimensional toy model. 
In general the energy is of the following form
\beq
E = E_{BPS} (\Phi) + \zeta V(\Phi) \ ,
\eeq
with the property that $E_{BPS} (\Phi)$ has a flat direction in a subspace $\Phi_{BPS} (\lambda)$  while   $ V(\Phi)$ is a generic potential which lifts this degeneracy  and $\zeta$ is a parameter which we want to send to zero. 
 The BPS  property implies that, as we perform any expansion around a BPS solution
\beq
\label{expansionaroundbps}
\Phi =  \Phi_{BPS} ({\lambda}) +\Delta \Phi_{\perp BPS} + \Delta \Phi_{\parallel  BPS} \ ,
\eeq
the energy is sensitive only to the  fluctuations in the space perpendicular to the moduli space $ \delta \Phi_{\perp BPS}(x)$ : 
\beq
\label{bpsproperty}
E_{BPS}(\Phi) = E_{BPS\, bound}   + \frac{1}{2}  \frac{ \partial^2 E_{BPS}(\Phi)  }{\partial {\Phi_{\perp}}^2 }  \Delta \Phi_{\perp BPS}^2 + \dots \ .
\eeq
Note that the fluctuations $\Delta_{\perp}$ and $\Delta_{\parallel}$ live in a vector space but for simplicity we avoid  writing explicitly  the vectorial indices.

To prove our claim we want to determine the minimum of this expression for small $\zeta$.  
We use the expansion (\ref{expansionaroundbps}) around a generic point  in the BPS moduli space. Our derivation will provide at the end the correct value  $\Phi_{BPS} ({\lambda})$ to which the solution is flowing as $\zeta \to 0$.

It is convenient to separate the fluctuations into two parts
\bea
\label{doubleexpansion}
&& \Delta \Phi_{\perp BPS} =\bar{\delta} \Phi_{\perp BPS}+{\delta} \Phi_{\perp BPS} \ ,\nonumber \\
&& \Delta \Phi_{\parallel  BPS} = \bar{\delta} \Phi_{\perp BPS}+{\delta} \Phi_{\perp BPS} \ ,
\eea
where $\bar{\delta}$ is the fluctuation of the solution around the $\zeta \to 0$ limit, while ${\delta}$ describes any other fluctuation which we may consider when we try to minimize the energy. 
The total energy expansion, up to the second order, is then given by
\beq
\label{energytotalfluctuation}
E&=&  E_{BPS\, bound}   + \frac{1}{2}  \frac{ \partial^2 E_{BPS}(\Phi)  }{\partial {\Phi_{\perp}}^2 }  \Delta \Phi_{\perp BPS}^2 \nonumber \\
&& + \zeta \left( V( \Phi_{BPS} ({\lambda})) + \frac{\partial V(\Phi) }{\partial {\Phi_{\perp}}} \Delta \Phi_{\perp BPS}  + \frac{\partial V(\Phi) }{\partial {\Phi_{\parallel}}} \Delta \Phi_{\parallel BPS} \right. \nonumber \\
&& \left.\frac{1}{2}\frac{\partial^2 V(\Phi) }{\partial {\Phi_{\perp}}^2} \Delta \Phi_{\perp BPS}^2   +  \frac{\partial^2 V(\Phi) }{\partial {\Phi_{\parallel}}\partial {\Phi_{\perp}}} \Delta \Phi_{\perp BPS}  \Delta \Phi_{\parallel BPS} +   \frac{1}{2}\frac{\partial^2 V(\Phi) }{\partial {\Phi_{\parallel}}^2} \Delta \Phi_{\parallel BPS}^2 \right) \ , \nonumber \\
\eeq
where $\Delta$'s are given by (\ref{doubleexpansion}).

We first evaluate the perpendicular part of the fluctuation $ \bar{\delta} \Phi_{\perp BPS}$. 
For this we have to set to zero the  term in (\ref{energytotalfluctuation}) proportional to  $ \delta \Phi_{\perp BPS}$: 
\beq  
 \left( \frac{ \partial^2 E_{BPS}(\Phi)  }{\partial {\Phi_{\perp}}^2 }  \bar{\delta} \Phi_{\perp BPS}   +  \zeta \frac{\partial V(\Phi) }{\partial {\Phi_{\perp}}} \right) \delta \Phi_{\perp BPS} =0 \ .
\eeq
Thus we have
\beq
\label{phiperp}
   \bar{\delta} \Phi_{\perp BPS} = -  \zeta \left(
 \frac{ \partial^2 E_{BPS}(\Phi)  }{\partial {\Phi_{\perp}}^2 } \right)^{-1} \frac{\partial V(\Phi) }{\partial {\Phi_{\perp}}} 
\eeq
and so we see that $\bar{\delta} \Phi_{\perp BPS}$ goes to zero linearly in $\zeta$.

Then we find the right value of ${\lambda}$  to which the solution flows as $\zeta \to 0$ and also the fluctuation  $ \bar{\delta} \Phi_{\parallel BPS}$.  This time is  the  term in (\ref{energytotalfluctuation}) proportional to  $ \delta \Phi_{\parallel BPS}$ that  must be set to zero: 
\beq
\label{derivativetan}
\zeta \left( \frac{\partial V(\Phi) }{\partial {\Phi_{\parallel}}}+\frac{\partial^2 V(\Phi) }{\partial {\Phi_{\parallel}}\partial {\Phi_{\perp}}} \bar{\delta} \Phi_{\perp BPS} +\frac{\partial^2 V(\Phi) }{\partial {\Phi_{\parallel}}^2} \bar{\delta} \Phi_{\parallel BPS}  \right) \delta \Phi_{\parallel BPS} = 0 \ .
\eeq
The leading term must be set to zero separately, and this gives
\beq
 \frac{\partial V(\Phi) }{\partial {\Phi_{\parallel}}} =0 \ .
\eeq
This, as anticipated before,  is the condition that determines the correct point of the BPS moduli space.  
Setting to zero the higher order terms in (\ref{derivativetan}) we get the fluctuation in the parallel direction
\beq 
\label{phipara}
\bar{\delta} \Phi_{\parallel BPS} = - \left(\frac{\partial^2 V(\Phi) }{\partial {\Phi_{\parallel}}^2} \right)^{-1}  \frac{\partial^2 V(\Phi) }{\partial {\Phi_{\parallel}}\partial {\Phi_{\perp}}}  \bar{\delta} \Phi_{\perp BPS},  
\eeq
where $ \bar{\delta} \Phi_{\perp BPS}$ is given in (\ref{phiperp}).
So  $\bar{\delta} \Phi_{\parallel BPS}$ also goes to zero linearly in $\zeta$.

So we note that the energy evaluated on the solution has the following expansion in $\zeta$:
\beq
E=  E_{BPS\, bound}   + \zeta \,  V( \Phi_{BPS} ({\lambda})) + {\cal O}(\zeta^2) \ .
\eeq
All the terms in this expression that depend on the fluctuations $ \bar{\delta} \Phi_{\perp BPS}$ and $\bar{\delta} \Phi_{\parallel BPS}$ are at least of order $\zeta^2$.

We give an illustrative example which supports these claims. It involves a two dimensional $(x,y)$ model with
\beq
E_{BPS} = x^2\ , \quad  \qquad V = y^2 + \alpha x + \beta x^2 + \gamma x y \ .
\eeq
The moduli space in this case is the line $x = 0$, so $\Phi_{\perp}$ corresponds to $x$ and $\Phi_{\parallel}$ to $y$.  The minimum of $ E_{BPS} (\Phi) + \zeta V(\Phi) $ can be computed exactly in this case and it corresponds to 
\beq
x =  - \frac{\alpha \zeta}{2 + 2 \beta \zeta - \gamma^2 \zeta/2 } \ , 
\qquad  \qquad 
y   =  \frac{\alpha \gamma \zeta}{4 + 4 \beta \zeta + \gamma^2 \zeta }  \ .
\eeq
As $\zeta \to 0$ this minimum flows to the point $(x,y) = (0,0)$ which is exactly the minimum of $V$ restricted to the line $x =0$. Moreover the perpendicular and parallel fluctuations as $\zeta \to 0$ are exactly the ones given  by  $(\ref{phiperp})$ and $(\ref{phipara})$, namely: 
\beq
&& \bar{\delta} x =  - \zeta\left(
 \frac{ \partial^2 E_{BPS}  }{\partial x ^2 } \right)^{-1} \frac{\partial V }{\partial x} =   - \frac{\alpha \zeta}{2 }  \ ,
\nonumber \\
&& \bar{\delta} y   = - \left(\frac{\partial^2 V }{\partial y^2} \right)^{-1}  \frac{\partial^2 V }{\partial x \partial y}  \, \bar{\delta} x =    \frac{\alpha \gamma \zeta}{4 }  \ .
\eeq

\section{The supersymmetric  baby Skyrme model}
\label{firstsusy}

In this section we consider various types of supersymmetric extensions of the baby Skyrme model. We use the conventions of \cite{Gates:1983nr} for ${\cal N}=1$ supersymmetry in $(2+1)$ dimensions, apart from the metric signature which we take $\eta^{\mu\nu}={\rm diag} (1,-1,-1)$. We will follow closely the supersymmetric constructions of Refs. \cite{Adam:2011hj,Adam:2013awa},  but with the inclusion  of some important extra terms.

First of all, let us say a few words about our notation.
A ${\cal N}=1$ superfield in $(2+1)$ dimensions has the following expansion in Grassmannian coordinates
\beq
\label{superfield1}
U = u + \theta^{\alpha} \psi_{\alpha} - \theta^2 F,
\eeq
where $\theta^{\alpha}$ is a Majorana spinor.
The tensors for raising and lowering the spinorial indices are $
C_{\alpha \beta} = \sigma_2 = -C^{\alpha \beta}$.
The covariant derivative which commutes with the supersymmetry generators is given by:
\beq
\label{covd1}
D_{\alpha} = \partial_{\alpha} + i \gamma^{\mu \beta}_{\alpha} \theta_{\beta} \partial_{\mu} \ ,
\eeq
and the gamma matrices are of the purely imaginary form:
\beq
\gamma^0=\sigma_2, \quad \gamma^1= i \sigma^3, \quad \gamma^2=i \sigma_1 .
\eeq

We consider the following terms in the Lagrangian density, of which we will write down explicitly  only their bosonic terms in the action.
The first term is the quadratic derivative term:
\beq
{\cal L}_2 = - \int d^2 \theta  g(U,\bar{U}) D^{\alpha} \bar{U} D_{\alpha} U = g(u,\bar{u}) \left(|F|^2 + \partial_{\mu} \bar{u} \partial^{\mu} u \right)   + {\rm ferm}  \ .
\eeq
Then we have five different higher derivative terms.
The first three of them are generated by considering a superfield of the following form
\beq
D_{\alpha} U  D_{\beta} \bar{U} D_{\xi} D_{\tau} U  D_{\rho} D_{\sigma} \bar{U}
\eeq 
with different contractions of the spinorial indices performed with the $C^{\alpha \beta}$ tensor. 
The three such  terms and their bosonic parts in the Lagrangian are given by:
\bea
\label{hd1}
{\cal L}_{4,1} &=& - \frac{1}{4}\int d^2  \theta h_1(U,\bar{U}) D_{\alpha} {U} D_{\alpha} \bar{U} D_{\beta} D_{\gamma} {U}  D^{\beta} D^{\gamma} \bar{U} \nonumber \\
&=&   h_1(u,\bar{u}) \left(|F|^4 + 2 |F|^2 \partial_{\mu} \bar{u} \partial^{\mu} u  + ( \partial_{\mu} \bar{u} \partial^{\mu} {u})^2 \right)   + {\rm ferm} \ ,
\eea
\bea
\label{extratermdue}
{\cal L}_{4,2} &=&  - \frac{1}{2}  \int d^2 \theta  h_2(U,\bar{U})  D_{\alpha} {U} D_{\beta} \bar{U} D^{\alpha} D^{\beta} {U}  D^{\gamma} D^{\gamma} \bar{U}  + {\rm h.c.}  \nonumber \\
&=& h_2(u,\bar{u}) \left(4|F|^4 +8|F|^2 \partial_{\mu} \bar{u} \partial^{\mu} u  -   F^2  \partial_{\mu} \bar{u} \partial^{\mu} \bar{u} - \bar{F}^2  \partial_{\mu} {u} \partial^{\mu} {u}\right)  + {\rm ferm} \ ,  \nonumber \\
\eea
\bea
\label{extratermtre}
{\cal L}_{4,3} &=&  -  \frac{1}{2}  \int d^2 \theta  h_3(U,\bar{U})  D_{\alpha} {U} D_{\beta} \bar{U} D^{\alpha} D_{\gamma} {U}  D^{\beta} D^{\gamma} \bar{U}   \nonumber \\
&=& h_3(u,\bar{u}) \left(|F|^4 + 2|F|^2 \partial_{\mu} \bar{u} \partial^{\mu} u +  |\partial_{\mu} {u} \partial^{\mu} u|^2 + \right. \nonumber \\
 & &\left. -  F^2  \partial_{\mu} \bar{u} \partial^{\mu} \bar{u}- \bar{F}^2  \partial_{\mu} {u} \partial^{\mu} {u}\right)  + {\rm ferm} \ . 
\eea

Note that the last two of these terms (${\cal L}_{4,2}$ and ${\cal L}_{4,3}$) were  not included in \cite{Adam:2011hj,Adam:2013awa} and they will be important in what follows.

The remaining  two higher derivative contributions are constructed from a superfield  of the form
\beq
D_{\alpha} U  D_{\beta} {U} D_{\xi} D_{\tau} \bar{U}  D_{\rho} D_{\sigma} \bar{U} + {\rm h.c.}
\eeq 
with different contractions of its spinorial indices. The  two terms that we need are:
 \bea
\label{extratermquattro}
{\cal L}_{4,4} &=&  -  \frac{1}{8}  \int d^2 \theta    h_4(U,\bar{U}) \left(  D_{\alpha} {U} D^{\alpha} {U} D_{\beta} D^{\beta} \bar{U}  D_{\gamma} D^{\gamma} \bar{U}    + {\rm h.c.} \right)\nonumber \\
&=& h_4(u,\bar{u}) \left(2|F|^4 +  F^2  \partial_{\mu} \bar{u} \partial^{\mu} \bar{u} + \bar{F}^2  \partial_{\mu} {u} \partial^{\mu} {u}\right) + {\rm ferm} \ , 
\eea
\bea
\label{extratermcinque}
{\cal L}_{4,5} &=&  -  \frac{1}{8}  \int d^2 \theta  h_5(U,\bar{U}) \left(    D_{\alpha} {U} D^{\alpha} {U} D_{\beta} D_{\gamma} \bar{U}  D^{\beta} D^{\gamma} \bar{U}          + {\rm h.c.}  \right) \nonumber \\
&=&h_5(u,\bar{u}) \left(|F|^4  + |\partial_{\mu} {u} \partial^{\mu} u|^2   + F^2  \partial_{\mu} \bar{u} \partial^{\mu} \bar{u} +  \bar{F}^2  \partial_{\mu} {u} \partial^{\mu} {u}\right) + {\rm ferm} \ . \nonumber \\
\eea

There are many other possible scalar superfield combinations which have the same number of superfields $U$ and same number of covariant derivatives $D_{\alpha}$. The previous list does not provide a complete classification.  But for our purposes, our choice of five terms is the minimal number we have to take into consideration. The reason for this is  the following. The bosonic sector of the higher derivative terms has five possible terms, which can be combined into a $5$-vector $B_i$:   
\beq
\label{vectornotation}
B_i = \left(|F|^4, \ |F|^2 \partial_{\mu} \bar{u} \partial^{\mu} u, \  (\partial_{\mu} \bar{u} \partial^{\mu} u)^2    , \ |\partial_{\mu} {u} \partial^{\mu} u|^2 , \ F^2  \partial_{\mu} \bar{u} \partial^{\mu} \bar{u} +{\rm h.c.} \right) 
\eeq
with $i=1,..5$.
A sum of the previous five terms in the Lagrangian, (\ref{hd1}), (\ref{extratermdue}), (\ref{extratermtre}), (\ref{extratermquattro}) and (\ref{extratermcinque}), gives  a generic linear combination of these terms in the bosonic sector
\beq
\sum_{i=1}^5
{\cal L}_{4,i} =  h_i(u,\bar{u}) M_{ij} B_j \ ,
\eeq
with the matrix $M_{ij}$  being 
\beq
M = \left(  \begin{array}{ccccc} 
1&2&1&0&0\\
4&8&0&0&-1\\
1&2&0&1&-1\\
2&0&0&0&1\\
1&0&0&1&1\\
\end{array}
\right) \ .
\eeq
Since the determinant of the matrix $M$ is different from zero,  the five terms are all linearly independent.
These five terms are then sufficient to construct any possible combination of such bosonic terms in the Lagrangian.
With the inclusion of more general higher derivative terms, we could  have a different fermionic sector with the same bosonic part. Exploring the full set of possibilities is beyond the scope  of this project. The four terms ${\cal L}_{4,i}$ with $i=2,3,4,5$ are the ones that arise in the ${\cal N}=2$ extended model, as we shall derive in Eq. (\ref{extendedcombination}), so it is natural to choose these. 
If we want to consider a generic ${\cal N}=1$ bosonic sector we need to add a fifth linearly independent one which is ${\cal L}_{4,1}$.

We also have a potential term with no derivatives
\beq
{\cal L}_0 = - \int d^2 \theta  w(U,\bar{U}) = \partial_u w F + \partial_{\bar u }w  \bar{F}  + {\rm ferm} \ .
\eeq
So the total lagrangian is  a sum of these various terms:
\beq
{\cal L} =
{\cal L}_2 + \sum_{i=1}^5
{\cal L}_{4,i} + {\cal L}_0 \ .  
\eeq

Next we look at the ways of recovering the baby Skyrme model in the bosonic sector. 
For this we want to sum all the bosonic terms,  integrate out the auxiliary field $F$,  and then constraint the remaining bosonic Lagrangian to be the one of the baby Skyrme model. 
The fermionic part of the Lagrangian contains purely fermionic terms and also the mixed ones. In fact, it is not possible to integrate out, in a closed form, the auxiliary field for the full Lagrangian, including the fermionic sector. 
So we do not have an on-shell form of the supersymmetric baby Skyrme model, with only the fields $u$ and $\psi$. 
Only for the solution in which  the fermions have been set to zero, which is always possible due to the form of their equation of motion, we recover the baby Skyrme model after integrating out the auxiliary field.

The most general bosonic baby Skyrme model is parametrized by three real and positive functions $V(u,\bar{u})$, $K(u,\bar{u})$, $S(u,\bar{u})$ and is of the form 
\bea
\label{anymodel}
{\cal L} =K(u,\bar{u})\partial_{\mu} \bar{u} \partial^{\mu} u +S(u,\bar{u}) \left( |\partial_{\mu} {u} \partial^{\mu} u|^2 - \partial_{\mu} \bar{u} \partial^{\mu} u \right) - V(u,\bar{u}) \ .
\eea
The specific cases considered in Section \ref{first} are
\beq
\label{specific}
 K =  \frac{1}{(1+|u|^2)^2} \ , \qquad  S = \frac{1}{(1+|u|^2)^4} \ ,   \qquad V = \frac{|u|^{2k}}{\left(1+|u|^2\right)^k}  \ .
\eeq

To proceed further we note that there are two different strategies to obtain the baby Skyrme Lagrangian; both strategies have been considered in the papers \cite{Adam:2011hj} and \cite{Adam:2013awa}. We will adopt the same strategies, but taking into consideration also the extra terms (\ref{extratermdue}) and (\ref{extratermtre}).

The first strategy in the one discussed in \cite{Adam:2011hj}. In it we try to combine the higher derivative terms in order to reproduce the baby Skyrme term with no additional auxiliary field terms.  
The baby Skyrme higher derivative term, in the notation of (\ref{vectornotation}), correspond to the vector $(0,0,-1,1,0)$. We then have the equation
\beq
S(u, \bar{u}) \left(0,0,-1,1,0 \right) = h_i(u, \bar{u})   M_{ij} \ ,
\eeq
which is solved by
\beq
h_i (u, \bar{u}) =  S(u, \bar{u})  \frac{1}{5} \left(-5,1,1,-2,4 \right) \ . 
\eeq
The Lagrangian in this case becomes  
\bea
\label{finallag} 
{\cal L} &=&  g |F|^2 +  g \partial_{\mu} \bar{u} \partial^{\mu} u+ \nonumber \\
&& + S(u, \bar{u})\left( |\partial_{\mu} {u} \partial^{\mu} u|^2 - \partial_{\mu} \bar{u} \partial^{\mu} u \right) +\nonumber \\
&& \partial_u w F + \partial_{\bar u }w  \bar{F}  + {\rm ferm} \ . 
\eea
After setting $\psi =  0 $ the auxiliary field can be solved by
\beq
\bar{F} = -\frac{ \partial_{u }w }{  g } \ ,
\eeq
and so the Lagrangian becomes 
\bea
\label{finallag2} 
{\cal L} &=&  g \partial_{\mu} \bar{u} \partial^{\mu} u -\frac{|\partial_{ u }w|^2}{g} \nonumber \\
&& + S(u, \bar{u})\left( |\partial_{\mu} {u} \partial^{\mu} u|^2 - \partial_{\mu} \bar{u} \partial^{\mu} u \right) \ . 
\eea

In this case we can then recover the baby Skyrmion theory by making the following choice
\bea
g(u,\bar{u}) &= & K(u,\bar{u}) \ ,\nonumber \\
|\partial_{ u }w|^2  &=& V(u,\bar{u}) K(u,\bar{u}) \ .
\eea
For example, for the specific choice (\ref{specific}), the solution for $w$ is given by the real integral
\beq
w(u, \bar{u}) = \int^{|u|} dx \frac{x^k}{(1+x^2)^{1+k/2}} \ .
\eeq
A solution of this form can be easily found whenever $V$ and $K$ are simply functions of $|u|$.

In the  second  approach (see \cite{Adam:2013awa}) we do not use any superpotential term, so we set ${\cal L}_0  =0$.  We then arrange the coefficients of the  higher derivative terms ${\cal L}_{4,i} $ so that only the terms proportional to $|F|^4$, $|F|^2$ or $|F|^0$ appear in the bosonic sector, {\it i.e.} we set to zero coefficients of the terms $F^2$ and $\bar{F}^2$.    In this case we  can integrate out explicitly the auxiliary field after we have set also $\psi =0$.   Finally, we arrange the coefficient of  the term with four time derivatives to vanish.

So for the terms proportional to $F^2$ and $\bar{F}^2$
to vanish we have 
\beq
 \left( - h_2 - h_3 + h_4 + h_5 \right) \left( F^2  \partial_{\mu} \bar{u} \partial^{\mu} \bar{u} + \bar{F}^2  \partial_{\mu} {u} \partial^{\mu} {u} \right) = 0
\eeq 
and so we require that
\beq
\label{h5s}
h_5=  h_2 + h_3 - h_4 \ .
\eeq
Then the  total Lagrangian becomes 
\bea
{\cal L} &=& g |F|^2 +  g \partial_{\mu} \bar{u} \partial^{\mu} u+ \nonumber \\
&&  + \left(h_1 + 5 h_2+  2 h_3 + h_4\right)|F|^4 + \nonumber \\
&& + \left(2h_1+ 8 h_2+ 2 h_3 \right)|F|^2 \partial_{\mu} \bar{u} \partial^{\mu} u +  \nonumber \\
&&+ h_1  (\partial_{\mu} \bar{u} \partial^{\mu} u)^2 +  \nonumber \\
&& + \left( h_2 +2 h_3 - h_4  \right) |\partial_{\mu} {u} \partial^{\mu} u|^2 + {\rm ferm} \ . 
\eea

Next we set $\psi =0$ and find that the auxiliary field can be solved to be given by
\beq
|F|^2 = -\frac{ g + \left(2 h_1+ 8 h_2+ 2 h_3 \right) \partial_{\mu} \bar{u} \partial^{\mu} u}{ 2 \left( h_1  + 5 h_2 + 2 h_3 + h_4 \right)} \ .
\eeq
Thus the bosonic Lagrangian, at this stage, becomes 
\bea
{\cal L} &=&  \frac{g \left(h_2+h_3+h_4\right)}{h_1+5 h_2+2 h_3+h_4} \partial_{\mu} \bar{u} \partial^{\mu} u +  \nonumber \\
&&+  \frac{h_1 \left(h_4-3 h_2\right)-\left(4 h_2+h_3\right){}^2}{h_1+5 h_2+2 h_3+h_4} (\partial_{\mu} \bar{u} \partial^{\mu} u)^2 +  \nonumber \\
&& + \left( h_2 +2 h_3 - h_4   \right) |\partial_{\mu} {u} \partial^{\mu} u|^2 + \nonumber \\
&&   -\frac{g^2}{4 \left(h_1+5 h_2+2 h_3+h_4\right)} \ .
\eea
Finally we have to impose the vanishing of coefficient of the terms with four time derivatives and this gives us the following constraint
\beq
-11 h_2^2-2 h_1 \left(h_2-h_3\right)+3 h_3^2+4 h_2 \left(h_3-h_4\right)-h_4^2 = 0 \ .
\label{relation}
\eeq

So the final bosonic Lagrangian of the baby Skyrme type is
\bea
\label{finallag3}
{\cal L} &=& \frac{g \left(h_2+h_3+h_4\right)}{h_1+5 h_2+2 h_3+h_4}\partial_{\mu} \bar{u} \partial^{\mu} u +   \nonumber \\
&& + \left( h_2 +2 h_3 - h_4\right) \left( |\partial_{\mu} {u} \partial^{\mu} u|^2 - \partial_{\mu} \bar{u} \partial^{\mu} u \right) +\nonumber \\
&& -\frac{g^2}{4 \left(h_1+5 h_2+2 h_3+h_4\right)} \ ,
\eea
with $h_{1,2,3,4}$ related by the condition (\ref{relation}).

If we want  a restricted baby Skyrme Lagrangian, which is (\ref{anymodel}) with $K=0$,  we have impose also the vanishing of the coefficient of the kinetic term in (\ref{finallag3}) and this  gives us
\beq
  h_2  = -   h_3  - h_4 \ . 
\eeq
The constraint (\ref{relation}) then becomes
\beq
\label{relationres}
2 \left(h_1-3h_3-4h_4\right) \left(2 h_3+h_4\right) = 0 \ .  
\eeq
We note that have two branches of solutions of this equation. The first branch $h_1-3h_3-4h_4 = 0$ gives an infinite potential, so we exclude it. The second one is
\beq
\label{h4s}
h_4 = -2 h_3 \ ,
\eeq
and gives us the following  bosonic Lagrangian
\beq
\label{lagrangianh}
{\cal L} =     5h_3   \left( |\partial_{\mu} {u} \partial^{\mu} u|^2 - \partial_{\mu} \bar{u} \partial^{\mu} u \right)- \frac{g^2}{4(h_1 + 5 h_3)} \ .  
\eeq

To match the restricted baby Skyrme model we can make the following choice:
\bea
h_3(u,\bar{u}) &=&  \frac{1}{5}  S(u,\bar{u})\ ,  \nonumber \\
h_1(u,\bar{u}) &=& \frac{\epsilon}{5}  S(u,\bar{u}) \ , \nonumber \\
g(u,\bar{u}) &=& 2 \sqrt{ \left(\frac{1}{5}+\epsilon \right) V(u,\bar{u}) S(u,\bar{u}) } \ ,
\label{branchtwo}
\eea
with $\epsilon > 1/5$.  This can also be rewritten using (\ref{h5s}) and (\ref{h4s}) as
\beq
\label{n2result}
h_i (u, \bar{u}) =  S(u, \bar{u})  \frac{1}{5} \left(\epsilon,1,1,-2,4 \right) \ .
\eeq
At the value $\epsilon = - 1/5$ we meet the first branch of solutions of (\ref{relationres}).
Not only we can then recover any restricted baby Skyrme model,  but  we also have a one parameter family labeled by $\epsilon$.

To recover the most general baby Skyrme model we need first to solve explicitly the constraint (\ref{relation}). We can express $h_1$ as a function of the others as follows
\beq
h_1=\frac{-11 h_2^2+4 h_2 h_3+3 h_3^2-4 h_2 h_4-h_4^2}{2 \left(h_2-h_3\right)} \ ,
\eeq
and then the bosonic Lagrangian becomes
\bea
\label{finallag4}
{\cal L} &=&\frac{2 g \left(h_3-h_2\right)}{h_2+h_3+h_4} \partial_{\mu} \bar{u}  \partial^{\mu} u +   \nonumber \\
&& (h_2+2 h_3-h_4)\left( |\partial_{\mu} {u} \partial^{\mu} u|^2 - \partial_{\mu} \bar{u} \partial^{\mu} u \right) +\nonumber \\
&& -\frac{g^2 \left(h_3-h_2\right)}{2 \left(h_2+h_3+h_4\right){}^2} \ .
\eea
We are thus left with having to solve the following three equations
\bea 
K &=&  \frac{g \left(h_3-h_2\right)}{ h_2+ h_3+h_4},\nonumber \\
S &=&  h_2 +2 h_3 - h_4,\nonumber \\
V &=&\frac{g^2 (h_3-h_2) }{2 \left(h_2 + h_3+h_4\right)^2}.
\label{fullequations}
\eea
One possible set of solutions  is  
\beq
g=\alpha \sqrt{V S}
\eeq
and
\bea
h_i &=& \left(\frac{K^2 +S V \left(\alpha ^2 - 4\right)-2 K \sqrt{S V} \alpha}{4 V}, \frac{-3 K^2+ 8 S V+2 K \sqrt{S V} \alpha }{40 V}, \right. \nonumber \\
  && \ \ \   \frac{K^2+4 S V+K \sqrt{S V} \alpha }{20 V}, \frac{K^2-16 S V+6 K \sqrt{S V} \alpha }{40 V}, \nonumber \\
&& \ \ \ \left. \frac{-K^2+16 S V-K \sqrt{S V} \alpha }{20 V} \right) \ , 
\eea
with $\alpha > 0$. So we have a one-parameter family of models for any bosonic baby Skyrme model. When $K=0$ we recover  the solutions (\ref{branchtwo}) and (\ref{n2result}) with $\alpha = 2\sqrt{1/5 + \epsilon}$.

\section{${\cal N}=2$ supersymmetric extensions}
\label{secondsusy}

In $\N=2$, the superspace spinor is complex.   We can write it as a sum of real and imaginary components as
\beq
\Theta^{\alpha} = \theta^{\alpha} + i \delta^{\alpha} \ ,
\eeq
where $\theta$ and $\delta$ are two Majorana spinors.
In particular, $\theta$ is the one that in our conventions corresponds to the $\N=1$ supersymmetry of the previous section.

The $\N=2$ covariant derivatives are
\beq
{\cal D}_{\alpha} = \partial_{\alpha} + i \sigma^{\mu}_{\alpha \dot{\alpha}} \bar{\Theta}^{\dot{\alpha}} \partial_{\mu} \ , \quad \qquad 
\bar{{\cal D}}_{\dot{\alpha}} = -\partial_{\dot{\alpha}} - i \Theta^{\alpha} \sigma^{\mu}_{\alpha \dot{\alpha}}  \partial_{\mu} \ .
\eeq
To proceed further we decompose the $\N=2$ covariant derivatives into sums of the $\N=1$ ones:
\beq
{\cal D}_{\alpha} = D_{\alpha}^{(\theta)} + i D_{\alpha}^{(\delta)} \ , \quad \qquad 
\bar{{\cal D}}_{\dot{\alpha}} = - D_{\dot{\alpha}}^{(\theta)} + i D_{\dot{\alpha}}^{(\delta)} \ ,
\eeq
where $D_{\alpha}^{(\theta)}$ and $D_{\alpha}^{(\delta)}$ are the same as (\ref{covd1}), respectively, for $\theta$ and $\delta$.
 $\N=2$ chiral and anti-chiral superfields satisfy the constraints $\bar{\D}_{\dot{\alpha}} U_{\N=2}=0 $ and ${\D}_{{\alpha}} \bar{U}_{\N=2} =0 $ and, when  expanded into components, they become
\bea
U_{\N=2} &=& u + i \Theta \sigma^{\mu} \bar{\Theta} \partial_{\mu} u + \frac{1}{4} \Theta\Theta \bar{\Theta} \bar{\Theta} \Box u + \sqrt{2} \Theta \psi -\frac{i}{\sqrt{2}} \Theta \Theta \partial_{\mu} \psi \sigma^{\mu} \bar{\Theta} + \Theta \Theta F \ , \nonumber \\
\bar{U}_{\N=2} &=&  \bar{u} - i \Theta \sigma^{\mu} \bar{\Theta} \partial_{\mu} \bar{u} + \frac{1}{4} \Theta\Theta \bar{\Theta} \bar{\Theta} \Box \bar{u} + \sqrt{2} \bar{\Theta} \bar{\psi} + \frac{i}{\sqrt{2}} \bar{\Theta} \bar{\Theta} {\Theta}  \sigma^{\mu} \partial_{\mu} \bar{\psi}+ \bar{\Theta} \bar{\Theta} \bar{F} \ . \nonumber \\
 \eea

When the ${\cal N}=2$ superfields are chiral or anti-chiral,  the following relations between the $\N=1$ covariant derivatives are satisfied:
\beq
 D_{\alpha}^{(\theta)}U_{\N=2} = 
 iD_{\alpha}^{(\delta)}U_{\N=2} \ , \quad \qquad  
 D_{\alpha}^{(\theta)}\bar{U}_{\N=2} = 
 - i D_{\alpha}^{(\delta)}\bar{U}_{\N=2} \ . 
\eeq
So all derivatives can be expressed as a function of a unique derivative which we take to be $
 D_{\alpha}^{(\theta)}$.  From now on we will denote $
 D_{\alpha}^{(\theta)}$ simply as $
 D_{\alpha}$.

The $\N=2$ superfields can be  expanded in powers of $\delta$ as follows
\bea
U_{\N=2} &=& U + i \delta^{\alpha}D_{\alpha}U -\frac{1}{2}\delta^{\alpha}\delta_{\alpha} D^{\beta} D_{\beta} U \ ,\nonumber \\
\bar{U}_{\N=2} &=& \bar{U} - i \delta^{\alpha}D_{\alpha}\bar{U} -\frac{1}{2}\delta^{\alpha}\delta_{\alpha} D^{\beta} D_{\beta} \bar{U} \ ,
\eea
where $U$ is the $\N=1$ superfield, like the one  defined as in (\ref{superfield1}), but with a different normalization for the fermionic field
\beq
\label{superfield2}
U = u + \sqrt{2} \theta^{\alpha} \psi_{\alpha} - \theta^2 F \ .
\eeq
In this formulation the $\theta$ dependence is hidden inside the $\N=1$ superfields $U$ and $\bar{U}$.

\begin{figure}[h!]
\centerline{
\epsfxsize12cm \epsfbox{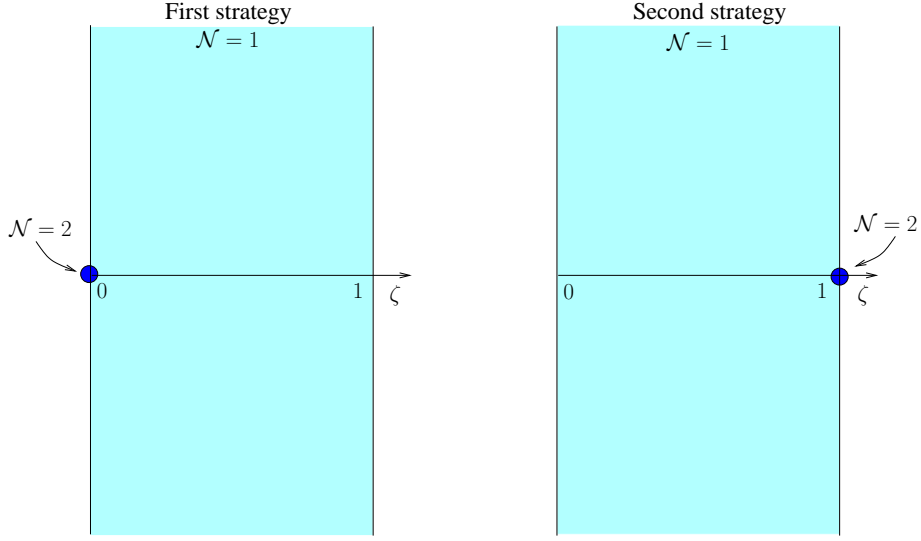}}
\caption{{\footnotesize Supersymmetric extensions of the baby Skyrme model. }}
\label{extensions}
\end{figure}

Returning to our problem,  we note that one $\N=2$ model is the pure sigma model arising from the Kahler potential, namely:
\bea
\label{kahlernod}
{\cal L}_2 &=& \int d^2 \Theta d^2 \bar{\Theta}\  {\cal K}(\bar{U}_{\N=2},U_{\N=2}) \nonumber \\
 &=& - \int d^2 \theta \ \bar{\partial} \partial {\cal K}(\bar{U},U)   D^{\alpha} \bar{U} D_{\alpha} U \ . 
\eea
When expressed in the $\N=1$ form, this shows that a $\N=1$ sigma model with a Kahler metric $ g(\bar{U},U) = \bar{\partial} \partial {\cal K}(\bar{U},U) $  has a hidden $\N=2$ supersymmetry  \cite{Witten:1977xn,DiVecchia:1977bs,D'Adda:1978kp,Zumino:1979et}. This model is one particular case of the theories arising from the first strategy of the previous section. We schematically describe these theories in Figure \ref{extensions}. The first strategy leads to the general  $\N=1$ extension of the baby Skyrme model. Any  theory with a parameter $\zeta$ defined in section \ref{first} can be extended to $\N=1$.  Among these theories, only one with $\delta = 0$ is extendable to $\N=2$, and this is only the case  if the metric is Kahler.

Another $\N=2$ extension is provided by the model discussed in \cite{Adam:2013awa} which is a particular extension of the restricted baby Skyrme model. So let us consider this model and expand it in the $\N=1$ formalism to see where it lies in the more general $\N=1$ extensions. 

This model is defined by
\beq
{\cal L} = 
{\cal L}_2 +  {\cal L}_4 \ ,
\eeq
where ${\cal L}_2$ is $(\ref{kahlernod})$ and 
${\cal L}_4$ is: 
\bea
\label{extendedcombination}
{\cal L}_4 &=& -\int d^2 \Theta d^2 \bar{\Theta}\  \frac{1}{10}{S}(\bar{U}_{\N=2},U_{\N=2})D^{\alpha}U_{\N=2}  D_{\alpha}U_{\N=2} D^{\alpha}\bar{U}_{\N=2}  D_{\alpha}\bar{U}_{\N=2} \nonumber \\
 &=& \int d^2 \theta \frac{1}{10} {S}(\bar{U},U) \left(  -  D_{\alpha} {U} D^{\alpha} {U} D_{\beta} D_{\gamma} \bar{U}  D^{\beta} D^{\gamma} \bar{U}          + {\rm h.c.} + \right. \nonumber \\
 && \ \   \qquad \qquad \qquad -  2   D_{\alpha} {U} D_{\beta} \bar{U} D^{\alpha} D_{\gamma} {U}  D^{\beta} D^{\gamma}  \bar{U}  +  \nonumber \\
 &&\ \  \qquad \qquad \qquad +  \frac{1}{2}D_{\alpha} {U} D^{\alpha} {U} D_{\beta} D^{\beta} \bar{U}  D_{\gamma} D^{\gamma} \bar{U}    + {\rm h.c.} +  \nonumber \\
 &&\ \ \qquad \qquad \qquad \left. -  D_{\alpha} {U} D_{\beta} \bar{U} D^{\alpha} D^{\beta} {U}  D^{\gamma} D^{\gamma} \bar{U}  + {\rm h.c.}  \right) + \dots \ .
\eea
For the last two lines we have performed an integration by parts of $D_{\alpha}$, and  the $\dots$ terms depend on derivatives of $S(\bar{U},U)$ but this do not affect the bosonic part of the Lagrangian. 
This expansion, taking into account also the coefficients in (\ref{extratermdue}), (\ref{extratermtre}), (\ref{extratermquattro}), (\ref{extratermcinque}), coincides exactly with (\ref{n2result}) for the choice $\epsilon =0$.

This other $\N=2$ extension belongs to the second strategy, as depicted in Figure \ref{extensions}. This
extension is also surrounded by other more general $\N=1$ extensions of the baby Skyrme model. Note that the two $\N=2$ extensions cannot be continuously connected by $\N=1$ extensions since they belong to two disconnected families.

\begin{figure}[h!]
\centerline{
\epsfxsize10cm \epsfbox{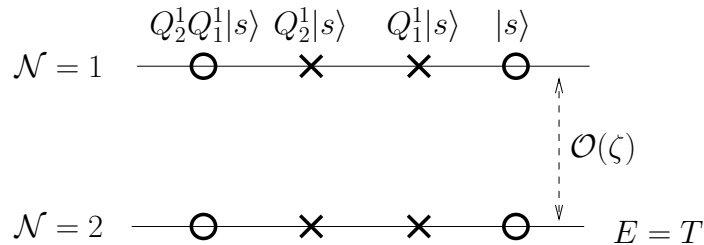}}
\caption{{\footnotesize A `short' $\N=2$ multiplet is lifted to a `long' $\N=1$ multiplet when supersymmetry is partially broken. }}
\label{multiplet}
\end{figure}

The extended supersymmetry algebra is usually  modified by the presence of  topological central charges \cite{Witten:1978mh}.
For the $\N=2$ $CP(1)$ sigma model in $(2+1)$ dimensions this charge has been computed explicitly  in \cite{Aoyama:1979jm,Ruback:1987sg}.  The algebra is given by
\bea
\{ Q^{I}_{\alpha}, Q^{J}_{\beta} \} =  \delta^{IJ} C_{\beta \rho} \gamma^{\mu\ \ \rho}_{\alpha} P_{\mu} + i \epsilon^{IJ} C_{\alpha \beta}T \ , 
\eea
where $Q^{I}_{\alpha}$ with $I=1,2$  are the two supersymmetry generators and $T$ is the topological charge. This quantum algebra has not yet been computed explicitly for the $\N=2$ theory corresponding to the restricted baby Skyrme model. 
Using the linear  combinations
\bea
Q_{\alpha} &=& \frac{1}{\sqrt{2}} (1+\gamma^2)_{\alpha}^{\beta} \left( Q^{1}_{\alpha} + i Q^{2}_{\alpha} \right) \ , \nonumber \\
\bar{Q}_{\alpha} &=& \frac{1}{\sqrt{2}} (1 - \gamma^2)_{\alpha}^{\beta} \left( Q^{1}_{\alpha} - i Q^{2}_{\alpha} \right) \ ,
\eea
and going into the soliton rest frame $P_{\mu} = (M,0,0)$, we have
\beq
\{ {Q}_{\alpha}, \bar{{Q}}_{\beta} \}
= 2 \left(
\begin{array}{cc}
M-T & 0 \\
0 & M + T
\end{array}\right)  \ .
\eeq
Solitons are the half-BPS states that annihilate the two supercharges $Q_1$ and $\bar{Q}_1$ and anti-solitons annihilate  instead $Q_2$ and $\bar{Q}_2$.
We can rewrite the four supercharges as ${\cal Q}_{1,2,3,4}$
\beq
{\cal Q}_1 = \frac{1}{\sqrt{2}} \left( Q^1_1 - Q^2_2 \right)\ ,\qquad 
{\cal Q}_2  = \frac{1}{\sqrt{2}} \left( Q^1_2 + Q^2_1 \right) \ ,\nonumber \\
{\cal Q}_3 = \frac{1}{\sqrt{2}} \left( Q^1_1 +  Q^2_2  \right)\ , \qquad 
{\cal Q}_4 =  \frac{1}{\sqrt{2}} \left( Q^1_2 - Q^2_1 \right) \ .
\eeq
The soliton annihilates ${\cal Q}_1$ and ${\cal Q}_2$ while the anti-soliton annihilates  
 ${\cal Q}_3$ and ${\cal Q}_4$· 
The supersymmetric multiplet is built around the bosonic soliton state $|s\rangle$ by acting with the broken supercharges: $|s\rangle$, $
{\cal Q}_3|s\rangle$, ${\cal Q}_4|s\rangle$, ${\cal Q}_3 {\cal Q}_4|s\rangle $. 
Since ${\cal Q}_1|s\rangle = {\cal Q}_2|s\rangle = 0$, we can equivalently write the multiplet as: $|s\rangle$,
 ${Q}^1_1|s\rangle$,  ${Q}^1_2|s\rangle$, ${Q}^1_1 {Q}^1_2|s\rangle $. When supersymmetry is broken to $\N=1$, the multiplet is simply lifted in a continuous way from the BPS bound. A short multiplet for $\N=2$ has in fact the same number of states of a long multiplet of $\N=1$ theory (see Figure \ref{multiplet}).

\section{Conclusions}
\label{conclusion}

In the first part of this paper we have laid ground for the near-BPS approximation, both analytically and numerically, using the baby Skyrme as a prototype model. Our analytical arguments also predict the rate of convergence to the BPS moduli space of solutions and, in particular, the rate of the deviation from the BPS moduli space (\ref{phiperp}) and (\ref{phipara}).  To test this rate of convergence we would need more powerful numerical methods that we have at our disposal at the moment. Also, a rigorous analytic proof would require more powerful functional analysis methods.  It would also be interesting to extend this analysis to the multi-soliton sector and to the bound states of  baby Skyrmions.  

We have also given a more complete construction of the $\N=1$ supersymmetric extensions of the baby Skyrme model,  generalizing the results of \cite{Adam:2011hj,Adam:2013awa}. 
Using two different strategies, we were able to construct two disconnected families of $\N=1$ theories, each of which possesses an $\N=2$ extension in which the solitons become BPS saturates. It has not, however, been possible, within the theories we have constructed, to construct a theory with a flow between these two $\N=2$ models without breaking all the supersymmetries. It is not clear, at present, whether a more general $\N=1$ framework exists which would allow such a continuous flow to be present. 



\section*{Acknowledgments} We thank P. Sutcliffe and A. Wereszczynski for useful discussions.  The  work of SB was initially funded by the EPSRC grant EP/K003453/1 and now by the grant `Rientro dei Cervelli, RLM' of the Italian government.

\end{document}